\documentclass[aps,pra,twocolumn,showpacs,preprintnumbers,amsmath,amssymb]{revtex4}
\usepackage{amssymb}
\usepackage{graphicx}
\usepackage{dcolumn}
\usepackage{bm}

\begin{document}

\input{epsf}

\title{Reentrance of Bose-Einstein condensation in spinor atomic gases in magnetic field }

\author{Chengjun Tao, Zhirong Jiao, and Qiang Gu}\email{qgu@sas.ustb.edu.cn}
\affiliation{Department of Physics, University of Science and
Technology Beijing, Beijing 100083, People's Republic of China}

\begin{abstract}
We calculate the Bose-Einstein condensation (BEC) temperature of
spin-1 atomic bosons in external magnetic field, taking into account
the influence of the quadratic Zeeman effect. In case that the
quadratic Zeeman coefficient is positive, the BEC temperature
exhibits a nontrivial dependance on the magnetic field and a
magnetic-field-induced reentrant phenomenon of BEC is observed. This
phenomenon could be well understood by the competition between the
linear and quadratic Zeeman effects. Reentrance of BEC in a trapped
spinor Bose gas is also discussed.

\end{abstract}

\date{\today}

\pacs{03.75.Mn, 03.75.Hh, 05.30.Jp, 75.30.Kz}

\maketitle

\section{introduction}
Since the first observation of spin-1 $^{23}$Na Bose-Einstein
condensation (BEC) in the optical trap \cite{Stamper,Stenger}, Bose
gas with spin degrees of freedom, called the spinor Bose gas
\cite{Ho,Ohmi}, becomes one of the central topics in cold atomic
physics. The spin degrees of freedom permit a variety of additional
exotic phenomena to be manifest. Considerable experimental and
theoretical works have been dedicated to investigating static or
dynamic properties relevant to spins of the condensate, including
magnetic ordering \cite{Ho,Ohmi,Gu1,Yi}, coherent spin-dynamics
\cite{Roma,Kron,Chang,Sch,Black}, spin-domains or textures
\cite{Iso,Zhang,Sad,Gu2,Miz,Mur} and so on. In addition, a number of
researchers have laid eyes on finite-temperature properties of
spinor bosons, especially on the BEC phenomenon itself
\cite{Yamada,Simkin,Gu3,Szirmai,Iso2,Zhang2}.

Bose-Einstein condensation in a free spin-1 Bose gas has ever been
studied by Yamada \cite{Yamada}, Simkin and Cohen \cite{Simkin}. The
obtained results suggest that the Bose gas is more sensitive to
external magnetic field than a Fermi gas and the BEC temperature
rises up with the magnetic field due to the Zeeman effect. If there
are ferromagnetic (FM) couplings between atoms, BEC occurs in a more
fascinating way because the FM coupling induces a new phase
transition, the FM transition \cite{Gu3}. Its critical temperature
is called the Curie point. It is already demonstrated that the Curie
point is never below the BEC temperature and both critical points
are increased by the FM coupling \cite{Gu3}. These conclusions have
also been confirmed by other groups \cite{Szirmai}. Moreover,
Bose-Einstein condensation in an optically trapped spinor Bose gas
has been elaborately studied \cite{Iso2,Zhang2}.

In the present study, we concentrate on the Bose-Einstein
condensation in an $F=1$ spinor Bose gas by considering quadratic
Zeeman effect (QZE). As Stenger et al. indicated \cite{Stenger}, the
Zeeman energy of a spin-1 Bose atom is given by
\begin{eqnarray}\label{eq1}
E_{ze}^\sigma = E_0  - ph\sigma  + qh^2 \sigma ^2,
\end{eqnarray}
where $\sigma$ refers to the spin-$z$ index of hyperfine state
$\left| {F=1,m_F=\sigma} \right\rangle$ ($\sigma= +1, 0, -1$), $h$
to the magnetic field, and $E_0$ to the Zeeman energy of the $m_F =
0$ state. The second term represents the linear Zeeman splitting and
the last term arises from the quadratic Zeeman effect. For a spinor
atom, the quadratic Zeeman effect can not be neglected due to the
hyperfine structure \cite{Schiff,Sant}. Very recently it is found
that the QZE can even be induced by the dipole trap, which is used
to confine cold atoms experimentally, for the spinor atom $^{52}{\rm
Cr}$ \cite{Sant}.

The QZE plays an important role for the understanding of spinor
atomic bosons, especially in strong magnetic field. Various of
effects caused by the quadratic Zeeman shift have been reported in
recent years. It, together with the linear Zeeman effect and
spin-dependant interactions, can not only lead to a number of novel
ground states \cite{Stenger,Sant}, but also affects significantly on
spin dynamics \cite{Roma,Kron} of the spinor condensates. The spin
dynamics is dramatically suppressed in magnetic field, owing to the
quadratic Zeeman splitting. The QZE can also influence on the vortex
state, and bring about new type vortex states in spinor condensates
\cite{Iso3}. In this paper, we aim at discussing the Bose-Einstein
condensation of the spin-1 Bose gas. Interestingly, we show that the
QZE can induce reentrance of BEC in the Bose gas.

The spin-1 Bose gas is described by the following Hamiltonian:
\begin{eqnarray}\label{eq2}
\mathcal {H} &=& \int d\textbf{r}\left[ \frac{{\hbar ^2 }}{{2m^*
}}\nabla \psi_{_a }^\dag   \cdot \nabla \psi_a + \frac{c_0} {2}
\psi_a^\dag  \psi_{b}^\dag \psi_{b} \psi_a \right.\nonumber\\
&&+ \frac{c_2}{2}\psi_a^\dag \psi_{a'}^\dag  \textbf{F}_{ab} \cdot
\textbf{F}_{a'b'} \psi_{b'} \psi_b \nonumber\\
&& \left. - ph\psi_a^\dag  F^z_{ab} \psi_b + qh^2 \psi_a^\dag
(F^z_{ab})^2 \psi_b \frac{}{}\right],
\end{eqnarray}
where $\psi_a (\textbf{r})$ is the field annihilation operator for
an atom in hyperfine state $\left| {1,a} \right\rangle$ at point
$\textbf{r}$. The first term represents kinetic energy. The terms
with coefficients $c_0$ and $c_2$ denote the spin-independent and
spin-dependent interactions, respectively. The forth and fifth terms
describe the linear and quadratic Zeeman effects, respectively. We
suppose that the linear Zeeman coefficient $p$ is always positive
hereinafter, but the quadratic Zeeman coefficient $q$ can be either
positive or negative. $\textbf{F}$ is a vector which consists of
three components of the $3\times3$ Pauli spin-1
matrices\cite{Ho,Ohmi}.

This paper is organized as follows. In Sec. II, we calculate the
condensation temperature and the condensate fraction of homogenous
spinor gases in the external magnetic field. The reentrance
phenomenon is observed in case of $q>0$ and a detail explanation is
given on how it is induced by the quadratic Zeeman effect in a free
gas. The Sec. III extends the study to the trapped interacting
gases. The interaction between atoms is treated by using a
mean-field approach. It is suggested that the interaction tends to
decrease the condensation temperature, but the reentrance phenomenon
sustains. The last section gives a brief summary.

\section{Reentrance in a homogenous gas}
First we consider a free spin-1 Bose gas, ignoring the interactions
in the Hamiltonian (\ref{eq2}). Then the effective Hamiltonian for
the grand canonical ensemble reads
\begin{eqnarray}\label{eq3}
{\hat H} - {\hat N}\mu  = \sum\limits_{k\sigma } {( \varepsilon_k  -
\mu - ph\sigma + qh^2 \sigma^2 ){\hat n}_{k\sigma } },
\end{eqnarray}
where $\varepsilon _k = \hbar ^2k^2 / 2m^\ast $ is the kinetic
energy of free particles with mass $m^\ast $, $\mu $ is the chemical
potential, and ${\hat N}$ is the operator of total particle number.
Since the Hamiltonian is diagonal, we may calculate the grand
thermodynamical potential
\begin{eqnarray}\label{eq4}
\Omega  =  - \frac{1}{\beta }\ln Z =  - \frac{1}{\beta }\ln Tr\left[
e^{ - \beta ({\hat H} - {\hat N}\mu)}\right],
\end{eqnarray}
where $Z = Tr \left[ e^{ - \beta ({\hat H} - {\hat N}\mu)}\right]$
is the partition function, and $\beta = 1 / k_B T$. The density of
particles is derived as
\begin{eqnarray}\label{eq5}
\bar n =  - \frac{1}{V}\left( {\frac{{\partial \Omega }}{{\partial
\mu _\sigma  }}} \right)_{T,V}  = \frac{1}{V}\sum\limits_{k\sigma }
{n_{k\sigma } },
\end{eqnarray}
where \emph{V} is the volume of the system, and $\bar {n} = N / V$
is the particle density. From Eq. (\ref{eq5}), we can obtain the
basic equation determining the phase diagram of the spin-1 Bose
system,
\begin{eqnarray}\label{eq6}
1 &=& n_c + \left(\frac{m^* k_B T}{2\pi\hbar ^2 \bar n^{2/3} }\right)^{3/2}
    \left [f_{3/2} \left( - \frac{\mu +ph - qh^2}{k_B T}\right) \right. \nonumber\\
  &&+ \left. f_{3/2} \left( -  \frac{\mu}{{k_B T}}\right)
    + f_{3/2} \left( - \frac{\mu - ph - qh^2}{k_B T}\right)\right],
\end{eqnarray}
where $n_c=\bar{n}_c/ \bar {n}$ is the condensate fraction, $\bar
{n}_c $ is the condensate density, and $f_s (x)$ is the
polylogarithm function defined as~\cite{Gu3}
\begin{eqnarray}\label{eq7}
f_s (x) \equiv Li_s (e^{ - x} ) = \sum\limits_{k = 1}^\infty
{\frac{{(e^{ - x} )^k }}{{k^s }}},~x\ge 0.
\end{eqnarray}

Above the condensation temperature $T_c $, the condensate fraction
$n_c=0$ and the chemical potential $\mu$ is determined by Eq.
(\ref{eq6}) for each given temperature. At $T_c $, atoms lying on
the lowest energy sub-level begin to condense. In order to determine
$T_c $, one have to make it clear which sub-level is the lowest
first. From Eq. (\ref{eq3}) it is easy to see that the energy level
for the ${m_{F}=1}$ state is lower than the other two as long as the
external magnetic field $h$ is sufficiently weak, so the $m_{F}=1$
atoms condense once the temperature is below $T_c $. The chemical
potential $\mu$ satisfies the equation ${\mu + ph - qh^2<0}$ above
$T_c$, and ${\mu + ph - qh^2=0}$ at and below $T_c$. Then the
condensate fraction is calculated by the following equation
\begin{eqnarray}\label{eq8}
1 &=& n_c+ \left(\frac T{T^0_c}\right)^{3/2}\left.\frac1{3\zeta(3/2)}
    \right[ \zeta(3/2) \nonumber\\
  &&+ \left. f_{3/2} \left(\frac{{ph-qh^2}}{{k_B T}}\right)
    + f_{3/2} \left(\frac{{2ph}}{{k_B T}}\right)\right],
\end{eqnarray}
where $$T^0_c  = \frac{2\pi\hbar^2}{m^*k_B} \left( \frac{\bar n}
{3\zeta(3/2)} \right)^{2/3}$$ is the BEC temperature for a free
spin-1 Bose gas without external field and $ \zeta(3/2) = f_{3/2}
(0)\approx 2.612$ is the Riemann zeta function.

As the external magnetic field becomes stronger, the energy level
structure may be modified because of the influence of the QZE,
especially in the case of $q>0$. When $qh^2>ph$, the $m_{F}=0$
sub-level becomes lower than the $m_{F}=1$ sub-level, and $m_{F}=0$
atoms condense in place of $m_{F}=1$ atoms correspondingly.
Therefore, $\mu\to 0$ as $T \to T_c$. Below $T_c$, Eq. (\ref{eq6})
reduces to
\begin{eqnarray}\label{eq9}
1 &=& n_c + \left(\frac T{T^0_c}\right)^{3/2}\frac1{3\zeta(3/2)}
   \left[ f_{3/2} \left( - \frac{{ph-qh^2}}{{k_B T}}\right) \right. \nonumber\\
 &&\left. + \zeta(3/2) + f_{3/2}\left(\frac{{ph+qh^2 }}{{k_B T}} \right)\right] .
\end{eqnarray}

\begin{figure}[tb]
\centerline{\includegraphics[width=0.4\textwidth,keepaspectratio=true]{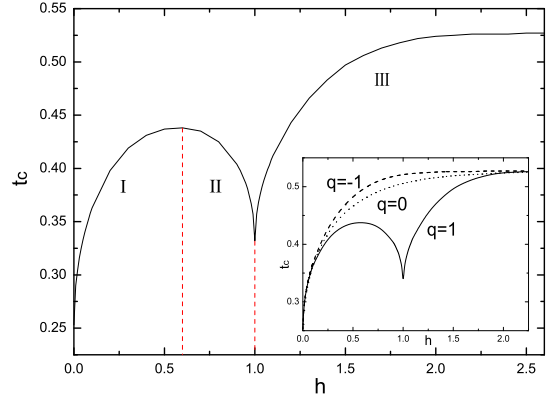}}
\caption{Reduced BEC temperature $t_c$ vs external magnetic field
$h$ for the spin-1 Bose gas with $q=1$. The inset plots BEC
temperatures for all the three cases of $q=-1, 0$ and $1$ so as to
make a comparison.}\label{fig1}
\end{figure}

Based on above discussions, we produce the $t_c$-$h$ phase diagram
numerically, as plotted in Fig. \ref{fig1}. In our calculations, a
reduced temperature $t = T/[T_c^0 ({3\zeta(3/2)})^{2/3}] =
({m^*k_BT})/({2\pi \hbar ^2\bar {n}^{2 / 3}})$ is introduced, and
then $ph$ and $qh^2$ must be re-scaled by the factor of $k_BT_c^0
({3\zeta(3/2)})^{2/3}$. The particle density is suppose to be $1$;
$p$, $q$ and $h$ are chosen to be dimensionless parameters and by
setting the re-scale factor to be $1$. For simplicity, we fix the
value of $p$ as $p=1$ and place emphasis on discussing three
distinct cases with respect to the QZE, $q=-1, 0$, and $1$. As shown
in the inset of Fig. \ref{fig1}, the phase diagram for the $q=-1$ or
$0$ case is very simple. The BEC temperature $t_c$ grows
monotonously as $h$ increases. In contrast, the $q=1$ case is very
intriguing: $t_c$ increase with $h$ first, and then decreases. After
dropping to a minimum value at $h=1.0$, it grows up again. During
the temperature range of $t\sim 0.33$ to $t\sim 0.43$, a typical
reentrant phenomenon of BEC induced by the magnetic field $h$ is
observed.

\begin{figure}[tb]
\includegraphics[width=0.4\textwidth,keepaspectratio=true]{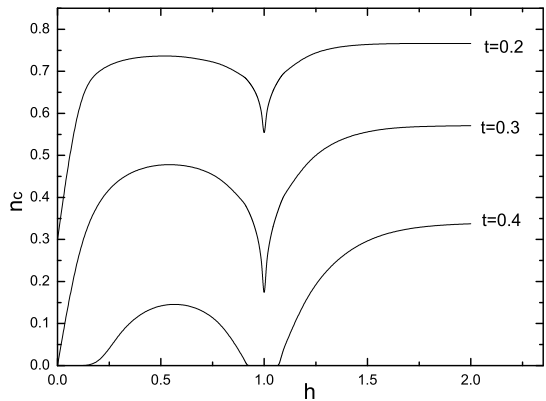}
\caption{Condensate fraction $n_c$ vs external magnetic field $h$ at
different temperatures $t=0.2, 0.3, 0.4$.}\label{fig2}
\end{figure}

To further demonstrate the reentrant phenomenon in spin-1 atomic
bosons with $q=1$, we calculate the condensate fraction $n_c$ at
some given temperatures, as shown in Fig. {\ref{fig2}}. The
condensed particles become dramatically less around $h=1.0$,
corresponding to the minimum point of $t_c$ in Fig. {\ref{fig1}}.
Especially at $t=0.4$, the condensate vanishes around $h=1.0$, and
then reappears in stronger field. The reentrant phenomenon can be
interpreted as the consequence of the competition between linear and
quadratic Zeeman effects, as discussed in the following.

\begin{figure}[tb]
\centerline{\includegraphics[width=0.4\textwidth,keepaspectratio=true]{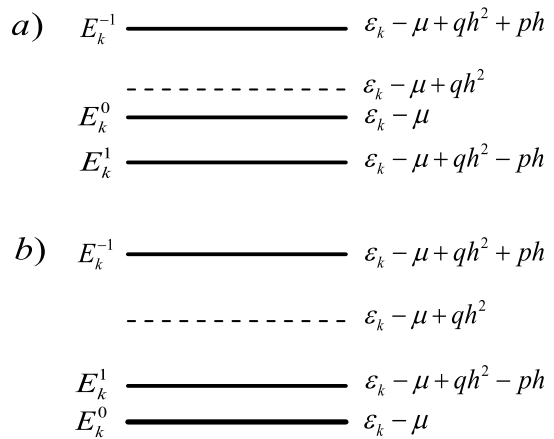}}
\caption{Energy splittings of the three spin components in case of
$q>0$ with $h$ being relatively weak (a) or strong (b).}\label{fig3}
\end{figure}

Without magnetic field, the energy spectra for the three hyperfine
spin states, $E_k^\sigma = \varepsilon_k - \mu - ph\sigma + qh^2
\sigma ^2$, are degenerate. Therefore, each spin state is equally
occupied, $\bar{n}_1 = \bar{n}_0 = \bar{n}_{-1} = \bar{n}/3$.
$\bar{n}/3$ is just the critical particle density, which determines
the BEC temperature to be $t^0_c=\{ 1/{[3\zeta(3/2)]}
\}^{2/3}\approx 0.2535$. When external magnetic field is applied,
hyperfine energy levels are shifted due to both the linear and
quadratic Zeeman effects. Fig. \ref{fig3} shows schematically the
splitting of energy levels for atoms with positive quadratic Zeeman
coefficient, $q>0$. The energy dispersion with respect to $k$ is not
depicted explicitly.

When $h$ is relatively week, the linear Zeeman effect dominates over
the QZE and the energy level splitting is shown in Fig.
\ref{fig3}(a), which satisfies $E_k^1<E_k^0<E_k^{-1}$. Consequently,
$\bar{n}_1 > \bar{n}_0 > \bar{n}_{-1}$, and $\bar{n}_1$ acts as the
critical density of particles. Because $\bar{n}_1 > \bar{n}/3$, the
BEC temperature is larger than $t^0_c$. As $h$ increases gradually,
the three energy levels become more separated so that $\bar{n}_1$
and thus $t_c$ tend to growing. This region of $h$ is labeled as
\uppercase\expandafter{\romannumeral1} in Fig. \ref{fig1}. When
$h=p/(2q)=1/2$, the separation between $E_k^1$ and $E_k^0$ arrives
at the largest value, $p^2/(4q)$. However, Fig. \ref{fig1} shows
that Region \uppercase\expandafter{\romannumeral1} terminates at a
field $h_1^*$ larger than $1/2$. During $1/2<h<h_1^*$, though the
separation between $E_k^1$ and $E_k^0$ becomes shrinking, the
separation between $E_k^1$ ($E_k^0$) and $E_k^{-1}$ is still
enlarged. As a result, more atoms occupy the $m_F=0$ and $m_F=1$ so
that $\bar{n}_1$ keeps increasing with $h$, until $h=h_1^*$.

In Region \uppercase\expandafter{\romannumeral2} of Fig. \ref{fig1},
the quadratic Zeeman term $qh^2$ becomes dominating, with the result
that $E_k^0$ becomes closer to the $E_k^1$ significantly. Although
$E_k^1$ is still the lowest energy level, the occupation number
$\bar{n}_1$ begins to decrease and $t_c$ drops accordingly. But
otherwise, $\bar{n}_0$ increases. Since $E_k^{-1}$ is far higher
than $E_k^0$ and $E_k^1$, few atoms occupy the $m_F=-1$ energy
level. When $h$ approaches a particular value, $h_2^*=p/q=1$, $E_k^0
= E_k^1$ and hence $\bar{n}_0 = \bar{n}_1$ which is slightly smaller
than $\bar{n}/2$. So the BEC temperature $t_c\approx 0.3327$ is
slightly higher than $(3/2)^{2/3} t_c^0\approx 0.3321$.

Increasing $h$ further, $E_k^0$ becomes the lowest energy level in
place of $E_k^1$, as indicated in Fig. \ref{fig3}(b). Therefore the
condensed atoms are in the $m_F=0$ state, instead of in the $m_F=1$
state. Meanwhile the BEC temperature $t_c$ grows up again. As the
stronger $h$ makes the three energy levels separate farther and
farther, $\bar{n}_0 \to \bar{n}$ and $t_c \to 3^{2/3} t_0$ in the
high field limit. This point is noted in Region
\uppercase\expandafter{\romannumeral3} of Fig. \ref{fig1}.

In case of ${q}=0$ or ${q}=-1$, $E_k^1$ remains the lowest energy
level all the way in magnetic field and only the $m_F=1$ atoms can
condense. The magnetic field $h$ persists in enhancing splittings of
the three energy levels. Correspondingly, the BEC temperature grows
monotonously with $h$ tending to stronger, as shown in the inset of
Fig. \ref{fig1}.

We note that the reentrant phenomenon has been discovered in a
number of quantum many-particle systems, especially in some
unconventional superconductors. For example, it is reported that
there appear successively two distinct superconducting phases with
the pressure in heavy fermion material CeCu$_2$Si$_2$~\cite{Yuan}. A
magnetic-field induced reentrance of superconductivity has ever been
observed in a pseudo-ternary Eu-Sn material~\cite{Meul}. For cold
atoms, Kleinert {\it et al}. have discussed reentrance in the
quantum phase transitions of an interacting Bose gas~\cite{Klein}.
They suggested that $T_c$ shifts upward with the interacting
strength first, then it is suppressed to zero.

\begin{figure}[tb]
\centerline{\includegraphics[width=0.4\textwidth,keepaspectratio=true]{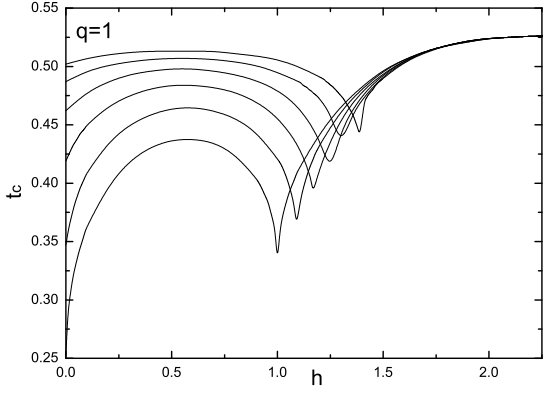}}
\caption{Reduced BEC temperature $t_c$ vs external magnetic field
$h$ for a ferromagnetically coupled spin-1 Bose gas. The coupling
$|c_2|=0, 0.2, 0.4, 0.6, 0.8,$ and $1$ from bottom to
top.}\label{fig4}
\end{figure}

The $^{87}$Rb gas happens to exhibit a positive QZE \cite{Sad} and
therefore it may be an appropriate example where the reentrant
phenomenon is expected to be observed. There exist weak FM couplings
between $^{87}$Rb atoms, as described by the $c_2$ term in Eq.
(\ref{eq2}) with $c_2<0$. So we proceed to examine how the FM
coupling affects on the reentrance of BEC, by decoupling the $c_2$
term via mean-field theory \cite{Gu3},
\begin{eqnarray}\label{eq10}
H_{c_2} =  - |c_2| {m}\sum\limits_{k}({\hat n}_{k1}-{\hat n}_{k-1}),
\end{eqnarray}
where $m$ is the normalized magnetization defined as
$m=\sum_{k}\langle {\hat n}_{k1}-{\hat n}_{k-1}\rangle/N$. Below the
BEC temperature, it is given by
\begin{eqnarray}\label{eq11}
m &=& \left(\frac T{T^0_c}\right)^{3/2}\frac1{3\zeta(3/2)}
   \left[ \zeta(3/2) - f_{3/2}\left(\frac{2ph + 2|c_2|m}{k_B T}\right) \right]\nonumber\\
   &&+n_c
\end{eqnarray}
if condensed atoms are in the $m_F=1$ state or
\begin{eqnarray}\label{eq12}
m &=& \left(\frac T{T^0_c}\right)^{3/2}\frac 1{3\zeta(3/2)}
   \left[ f_{3/2} \left(-\frac{{ph + |c_2|m - qh^2 }}{{k_B T}}\right) \right. \nonumber\\
   &&\left. - f_{3/2} \left(\frac{{ph + |c_2|m + qh^2 }}{{k_B T}}\right) \right]
\end{eqnarray}
if condensed atoms in the $m_F=0$ state.

Based on Eq. (\ref{eq10}), when the $c_2$ term is included, Eqs.
(\ref{eq6}), (\ref{eq8}) and (\ref{eq9}) should be rewritten by
replacing $ph$ with $ph+|c_2|m$. Hence it seems that the FM coupling
performs a role to enhance the linear Zeeman effect. Combining Eqs.
(\ref{eq8}) and (\ref{eq11}), or (\ref{eq9}) and (\ref{eq12}), we
obtain that the reentrant phenomenon still manifests, as seen in
Fig. 4. The FM coupling tends to stabilize the $m_F=1$ condensate,
so that the BEC temperature is upraised in the weak field region of
$h<h_2^*$ with the FM coupling increasing. Meanwhile, $h_2^*$ moves
to the high-field side.

\section{Reentrance in a trapped interacting Bose gas}
In experiments, atomic gases are usually confined in an effective
three-dimensional harmonic trap, so it is useful to extend the above
calculations to the trapped case. In this section, we consider a
system of interacting atoms trapped by the following potential
$$V(\textbf{r})=\frac 12 m^*\sum_i\omega_i^2r_i^2$$
where $\omega_i(i=x,y,z)$ denotes the angular frequency.

There are two interaction terms in the Hamiltonian (\ref{eq2}).
Actually, the spin-dependent interaction is rather weak in
comparison with the spin-independent interaction in cold atoms. For
example, in the $^{87}$Rb gas $c_2$ is about 2 orders of magnitude
smaller than $c_0$. Therefore, in the following calculations we
neglect the $c_2$ term and turn to deal with the $c_0$ term in some
detail.

For a trapped scalar Bose gas, the interaction can be well-treated
by the mean-field theory. The obtained results on the BEC
temperature and the condensate fraction are quite
accurate~\cite{SG,SG2}. The BEC temperature is shifted to lower
temperatures in the presence of repulsive interactions~\cite{SG}.
Hereinafter, we calculate the BEC temperature and condensate
fraction for spinor Bose gases in the magnetic field based on a
generalized mean-field approach from the scalar to the spinor Bose
gases~\cite{Zhang2}.

\begin{figure}[tb]
\centerline{\includegraphics[width=0.4\textwidth,keepaspectratio=true]{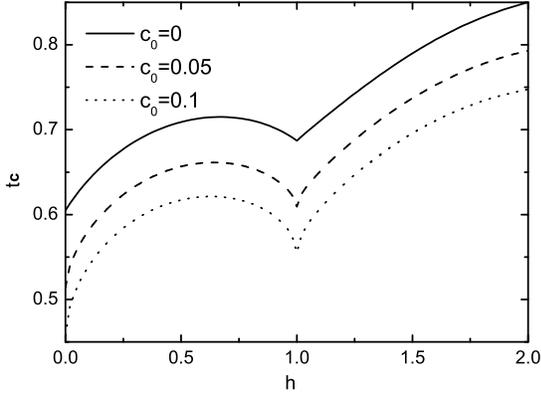}}
\caption{Reduced BEC temperature $t_c$ vs external magnetic field
$h$ for different interaction strength. The rescaled interaction
parameter $c_0^*=0, 0.05,$ and $0.1$ respectively.}\label{fig5}
\end{figure}

According to the mean-field theory~\cite{SG2}, the field operator
could be treated as $\psi_\sigma = \phi_\sigma+\delta\psi_\sigma$
with $\phi_\sigma = \langle\psi_\sigma\rangle$. Within the
Hartree-Fock (HF) approximation, the condensate can be described by
the finite-temperature generalization of the spinor Gross-Pitaevskii
equations (GPEs)~\cite{Zhang2}, which are given by
\begin{eqnarray}\label{eq13}
    i\hbar\frac{\partial}{\partial t}\phi_\sigma &=& \Big(-\frac{\hbar^2}{2m^*}\nabla^2+V(\textbf{r}) \nonumber \\
    &-& ph\sigma + qh^2 \sigma^2 + c_0(n+n_\sigma^T) \Big) \phi_\sigma
\end{eqnarray}
for the $m_F=\sigma=\pm 1$ condensation and
\begin{equation}\label{eq14}
    i\hbar\frac{\partial}{\partial t}\phi_0= \Big(-\frac{\hbar^2}{2m^*}\nabla^2+V(\textbf{r})+c_0(n+n_0^T) \Big) \phi_0
\end{equation}
for the $m_F=0$ condensation. Here $n_\sigma^c = |\phi_\sigma|^2$,
$n_\sigma^T = \langle\delta\psi_\sigma^\dag\delta\psi_\sigma\rangle$
and $n_\sigma = n^T_\sigma + n_\sigma^c$ describe the condensed,
noncondensed and total density of the $m_F=\sigma$ component,
respectively. $n=\sum_\sigma n_\sigma$ denotes the total density of
all atoms. As discussed in Sec. II, the $m_F=-1$ atoms will never
condense in the cases we considered. So for Eq. (\ref{eq13}), we
only discuss $\sigma=1$ case.

The dynamics of noncondensed atoms satisfy the following equations,
\begin{eqnarray}\label{eq15}
    i\hbar\frac{\partial}{\partial t}\delta\psi_1&=& \Big(-\frac{\hbar^2}{2m^*}\nabla^2+V(\textbf{r}) \nonumber \\
    &-&ph+qh^2+c_0(n+n_1) \Big) \delta\psi_1, \nonumber \\
    i\hbar\frac{\partial}{\partial t}\delta\psi_0&=& \Big(-\frac{\hbar^2}{2m^*}\nabla^2+V(\textbf{r})+c_0(n+n_0) \Big) \delta\psi_0, \nonumber \\
    i\hbar\frac{\partial}{\partial t}\delta\psi_{-1}&=& \Big(-\frac{\hbar^2}{2m^*}\nabla^2+V(\textbf{r}) \nonumber \\
    &+&ph+qh^2+c_0(n+n_{-1}) \Big) \delta\psi_{-1}.
\end{eqnarray}
Then we derive the effective energy spectra for all the three
components,
\begin{eqnarray}\label{eq16}
  \varepsilon_1(\textbf{k},\textbf{r})&=&\frac{\hbar^2k^2}{2m^*}+V(\textbf{r})-ph+qh^2+c_0(n+n_1), \nonumber \\
  \varepsilon_0(\textbf{k},\textbf{r})&=&\frac{\hbar^2k^2}{2m^*}+V(\textbf{r})+c_0(n+n_0),\\
  \varepsilon_{-1}(\textbf{k},\textbf{r})&=& \frac{\hbar^2k^2}{2m^*}+V(\textbf{r})+ph+qh^2+c_0(n+n_{-1}). \nonumber
\end{eqnarray}
Using the semiclassical approximation, the density of thermal atoms
with $m_F=\sigma$ becomes
\begin{eqnarray}\label{eq17}
  n^T_\sigma(\textbf{r}) &=& \int
  d^3\textbf{k}\left[{\rm exp}\left(\frac{\varepsilon_\sigma(k,\textbf{r})-\mu}{k_BT}\right)-1\right]^{-1},
\end{eqnarray}
where the chemical potential $\mu$ could be derived from Eq.
(\ref{eq13}) or (\ref{eq14}). On the basis of the Thomas-Fermi
approximation, one gets that
\begin{equation}\label{18a}
    \mu=V(\textbf{r})-ph+qh^2+c_0(n+n_1^T)
\end{equation}
when the linear Zeeman effect dominates and only the $m_F=\sigma=1$
atoms can condense, or
\begin{equation}\label{18b}
    \mu=V(\textbf{r}) + c_0(n+n_0^T)
\end{equation}
at sufficiently large field when the quadratic Zeeman effect
dominates and the $m_F=0$ atoms become condensed instead.

\begin{figure}[t]
\centerline{\includegraphics[width=0.4\textwidth,keepaspectratio=true]{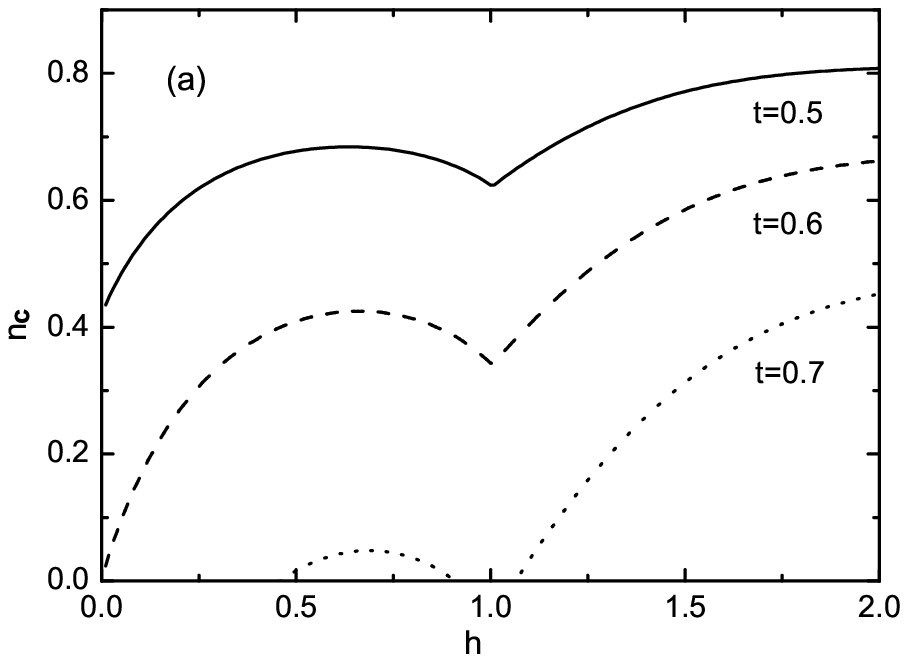}}
\centerline{\includegraphics[width=0.4\textwidth,keepaspectratio=true]{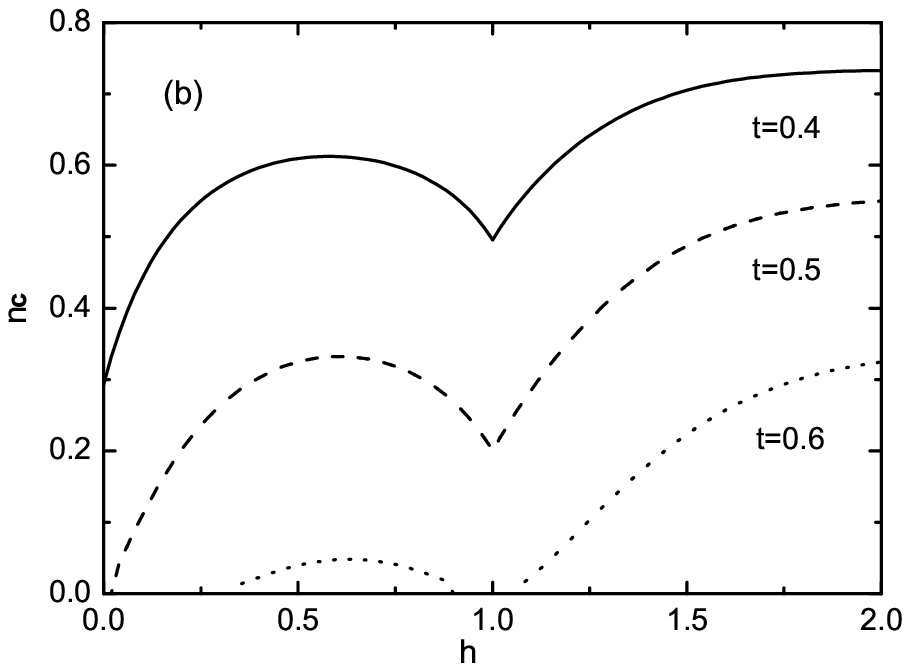}}
\caption{(a) Condensate fraction of noninteracting Bose gas in
harmonica trap at different temperatures $t=0.5, 0.6, 0.7$. (b)
Condensate fraction of interacting Bose gas in harmonic trap at
different temperatures $t=0.4, 0.5, 0.6$, for $c_0^*=0.1$
case.}\label{fig6}
\end{figure}

For simplicity, in our calculations all the energy parameters, such
as $k_BT$ and $c_0N$, are rescaled by a factor $k_BT^* = \hbar
[\sqrt{\pi/2} \omega_x\omega_y\omega_zN]^{1/3} =
[3\zeta(3)\sqrt{\pi/2}]^{1/3}k_BT_c^0$ where $T^0_c = \hbar
(\omega_x\omega_y\omega_z N)^{1/3}/[k_B (3\zeta(3))^{1/3}]$ is the
BEC temperature for $N$ trapped free spin-1 atoms without external
magnetic field. Then the reduced temperature reads $t=T/T^* =
[3\zeta(3)\sqrt{\pi/2}]^{-1/3}T/T_c^0$ and $t_c^0 =
[3\zeta(3)\sqrt{\pi/2}]^{-1/3}\approx0.6048$ for the $c_0=h=0$ case.

The  BEC temperature can be obtained self-consistently from Eqs.
(\ref{eq16}) and (\ref{eq17}), as shown in Fig. \ref{fig5}.
Obviously, the reentrance phenomenon sustains in the presence of the
trapping potentials. The $c_0$ term results in decrease of the BEC
temperature at all magnetic fields. As studied previously,
interactions tend to suppress the BEC in scalar bosons~\cite{SG} and
the BEC temperature drops linearly with the interaction strength
$c_0$ when $c_0$ is small enough. We estimate that the linear
relationship also holds in our case when both the resacled
interaction parameter $c_0^*$ and the resacled magnetic field $h$
are smaller than $0.01$. But in Fig. \ref{fig5}, $c_0^*$ is up to
$0.1$ and thus the linear relationship becomes invalid.

At last, we calculate the condensate densities below the critical
temperatures  according to Eq. (\ref{eq13}) (at low magnetic fields)
or Eq. (\ref{eq14}) (at high magnetic region). The densities of
noncondensed atoms can still been obtained from Eqs. (\ref{eq16})
and (\ref{eq17}). We get that the condensate mainly exists in the
center of the trap, while the noncondensed atoms mainly appear on
the edge region. The condensate fractions for both the $c_0^*=0$ and
$0.1$ cases are plotted in Fig. \ref{fig6}(a) and (b) respectively.
Again the reentrance phenomenon is illustrated for both cases. There
is no much more substantial changes in comparison with the
homogeneous case.

\section{Summary}
In summary, we have investigated the influence of the QZE on the BEC
temperature. In the case of free spinor Bose gas which exhibits a
positive QZE ($q>0$), the BEC temperature $T_c$ increases with the
external magnetic field $h$ first, and then begins decreasing until
dropping to a minimum value. $T_c$ increases again in stronger $h$.
Thus the system shows a reentrant phenomenon of BEC with respect $h$
in the phase diagram. Similar properties are also expected in
trapped interacting Bose gases. We suggest that this phenomenon can
be well understood by the competition between the linear and
quadratic Zeeman effects. We have also calculated the BEC
temperature in the case of negative QZE ($q<0$) and in the presence
of interactions.

{\bf Acknowledgement} This work is supported by the National Natural
Science Foundation of China (Grant No. 10504002), the Fok Yin-Tung
Education Foundation, China (Grant No. 101008), and the Ministry of
Education of China (Grant No. NCET-05-0098). We thank Dr. Yajiang
Hao for helpful discussions.


\begin{thebibliography}{}\label{sec:TeXbooks}
\bibitem{Stamper} D. M. Stamper-Kurn, M. R. Andrews, A. P. Chikkatur, S. Inouye,
H. J. Miesner, J. Stenger, and W. Ketterle, Phys. Rev. Lett. {\bf
80}, 2027 (1998).
\bibitem{Stenger} J. Stenger, S. Inouye, D. M. Stamper-Kurn, H.-J. Miesner, A.
P. Chikkatur, and W. Ketterle, Nature (London) {\bf 396}, 345 (1998).
\bibitem{Ho} T.-L. Ho, Phys. Rev. Lett. {\bf 81}, 742 (1998).
\bibitem{Ohmi} T. Ohmi and K. Machida, J. Phys. Soc. Jpn. {\bf 67}, 1822 (1998).
\bibitem{Gu1} Q. Gu, Phys. Rev. A {\bf 68}, 025601 (2003).
\bibitem{Yi} S. Yi, L. You, and H. Pu, Phys. Rev. Lett. {\bf 93}, 040403 (2004).
\bibitem{Roma}D. R. Romano and E. J. V. de Passos, Phys. Rev. A {\bf 70}, 043614 (2004).
\bibitem{Kron} J. Kronjager, C. Becker, M. Brinkmann, R. Walser, P. Navez,
K. Bongs, and K. Sengstock, Phys. Rev. A {\bf 72}, 063619 (2005).
\bibitem{Chang} M. S. Chang, C. D. Hamley, M. D. Barrett, J. A. Sauer, K. M.
Fortier, W. Zhang, L. You, and M. S. Chapman, Phys. Rev. Lett. {\bf
92}, 140403 (2004); M.-S. Chang, Q. Qin, W. Zhang, L. You, and M. S.
Chapman, Nat. Phys. {\bf 1}, 111 (2005).
\bibitem{Sch} H. Schmaljohann, M. Erhard, J. Kronjager, M. Kottke, S. van
Staa, L. Cacciapuoti, J. J. Arlt, K. Bongs, and K. Sengstock, Phys.
Rev. Lett. {\bf 92}, 040402 (2004).
\bibitem{Black} A. T. Black, E. Gomez, L. D. Turner, S. Jung, and P. D. Lett, Phys. Rev. Lett. {\bf 99}, 070403 (2007).
\bibitem{Iso}  T. Isoshima, K. Machida, and T. Ohmi, Phys. Rev. A {\bf 60}, 4857 (1999).
\bibitem{Zhang} W. Zhang, D. L. Zhou, M.-S. Chang, M. S. Chapman, and L.
You, Phys. Rev. Lett. {\bf 95}, 180403 (2005).
\bibitem{Sad} L.E. Sadler, J. M. Higbie, S. R. Leslie, M. Vengalattore, and
D. M. Stamper-Kurn, Nature {\bf 443}, 312 (2006).
\bibitem{Gu2} Q. Gu and H. Qiu, Phys. Rev. Lett. {\bf 98}, 200401 (2007).
\bibitem{Miz} T. Mizushima, N. Kobayashi, and K. Machida, Phys. Rev. A {\bf 70}, 043613 (2004).
\bibitem{Mur} K. Murata, H. Saito, and M. Ueda, Phys. Rev. A {\bf 75}, 013607 (2007).
\bibitem{Yamada} K. Yamada, Prog. Theor. Phys. {\bf 76}, 443 (1982).
\bibitem{Simkin} M. V. Simkin and E. G. D. Cohen, Phys. Rev. A {\bf 59}, 1528(1999).
\bibitem{Gu3} Q. Gu and R. A. Klemm, Phys. Rev. A {\bf 68}, 031604(R) (2003);
Q. Gu, K. Bongs, and K. Sengstock, {\it ibid.} {\bf 70}, 063609
(2004); C. Tao, P. Wang, J. Qin, and Q. Gu, Phys. Rev. B {\bf 78},
134403 (2008).
\bibitem{Szirmai}  K. Kis-Szabo, P. Szepfalusy, and G. Szirmai, Phys. Rev. A {\bf 72},
023617 (2005); G. Szirmai, K. Kis-Szabo, and P. Szepfalusy, Eur.
Phys. J. D {\bf 36}, 281 (2005); S. Ashhab, J. Low Tempt. Phys. {\bf 140} 51 (2005).
\bibitem{Iso2} T. Isoshima, T. Ohmi, and K. Machida, J. Phys. Soc. Jpn. {\bf 69}, 3864 (2000).
\bibitem{Zhang2}W. Zhang, S. Yi, and L. You, Phys. Rev. A {\bf 70}, 043611 (2004).
\bibitem{Schiff} L. I. Schiff and H. Snyder, Phys. Rev. {\bf 55}, 59 (1939).
\bibitem{Sant} L. Santos, M. Fattori, J. Stuhler, and T. Pfau, Phys. Rev. A {\bf 75}, 053606 (2007).
\bibitem{Iso3} T. Isoshima and S. Yip, J. Phys. Soc. Jpn. {\bf 75}, 074605 (2006).
\bibitem{Yuan} H. Q. Yuan, F. M. Grosche, M. Deppe, C. Geibel, G. Sparn, and F. Steglich, Science {\bf 302}, 2104 (2003).
\bibitem{Meul} H. W. Meul, C. Rossel, M. Decroux, and O. Fischer, G. Remenyi, and A. Briggs, Phys. Rev. Lett. {\bf 53}, 497 (1984).
\bibitem{Klein} H. Kleinert, S. Schmidt, and A. Pelster, Phys. Rev. Lett. {\bf 93}, 160402
(2004).
\bibitem{SG} S. Giorgini, L. P. Pitaevskii, and S. Stringari, J. Low Temp. Phys. {\bf 109}, 309 (1997); Phys. Rev. A {\bf 54}, R4633
(1996).
\bibitem{SG2}L. Pitaevskii and S. Stringari, \emph{Bose-Einstein Condensation} (Clarendon Press, Oxford, 2003), Chap 10.
\end{thebibliography}
\end{document}